# Analysis of ground penetrating radar data from the tunnel beneath the Temple of the Feathered Serpent in Teotihuacan, Mexico, using new multi-cross algorithms

Flor López-Rodríguez[a], Víctor M. Velasco-Herrera[b]*, Román Álvarez-Béjar[c], Sergio Gómez-Chávez[d], Julie Gazzola[d]

[a] Facultad de Ingeniería, Universidad Nacional Autónoma de México, Ciudad Universitaria, C.P. 04510, México

[b] Instituto de Geofísica, Universidad Nacional Autónoma de México, Ciudad Universitaria, C.P. 04510, México

[c] Instituto de Investigaciones en Matemáticas Aplicadas y en Sistemas, Universidad Nacional Autónoma de México, Ciudad Universitaria, México DF, México

[d] Dirección de Estudios Arqueológicos, Instituto Nacional de Antropología e Historia, Delegación Cuauhtémoc, C.P. 06060 México D.F., México

* Corresponding author

E-mail addresses:
FLR: florlopezr@yahoo.com.mx
VMVH: vmv@geofisica.unam.mx

# Abstract

The ground penetrating radar (GPR) —a non-invasive method based on the emission of electromagnetic waves and the reception of their reflections at the dielectric constant and electrical conductivity discontinuities of the materials surveyed— may be applied instead of the destructive and invasive methods used to find water in celestial bodies. As multichannel equipment is increasingly used, we developed two algorithms for multivariable wavelet analysis of GPR signals —multi-cross wavelet (MCW) and Fourier multi-cross function (FMC)— and applied them to analyze raw GPR traces of archeological subsurface strata. The traces were from the tunnel located beneath the Temple of the Feathered Serpent (The Citadel, Teotihuacan, Mexico), believed to represent the underworld, an outstanding region of the Mesoamerican mythology, home of telluric forces emanating from deities, where life was constantly created and recreated. GPR profiles obtained with 100 MHz antennas suggested the tunnel is 12-14 meters deep and 100-120 meters long with three chambers at its end, interpretations that were confirmed by excavations in 2014. Archeologists believe that due to the tunnel's sacredness and importance, one of the chambers may be the tomb of a ruler of the ancient city. The MCW and FMC algorithms determined the periods of subsurface strata of the tunnel. GPR traces inside-and-outside the tunnel/chamber, outside the tunnel/chamber and inside the tunnel/chamber analyzed with the MCW and filtered FMC algorithms determined the periods of the tunnel and chamber fillings, clay and matrix (limestone-clay compound). The tunnel filling period obtained by MCW analysis (14.37 ns) reflects the mixed limestone-clay compound of this stratum since its value is close to that of the period of the matrix (15.22 ns); periods of the chamber filling (11.40±0.40 ns) and the matrix (11.40±1.00 ns) were almost identical. FMC analysis of the tunnel obtained a period (5.08±1.08 ns) close to that of the chamber (4.27±0.82 ns), suggesting the tunnel and chamber are filled with similar materials. The use of both algorithms allows a deeper analysis since the similarities of the tunnel and chamber filling periods could not have been determined with the MCW algorithm alone. The successful application of the new multi-cross algorithms to archeological GPR data suggests they may also be used to search water and other resources in celestial bodies.

# 1 Introduction

Although spaceborne and airborne radars are widely used for remote sensing of surface and subsurface parameter estimation, ecological and agricultural monitoring and other purposes, the quality of primary radar images often does not provide the accuracy of data required for reliable data interpretation. To address this problem we have developed two algorithms for multivariable wavelet analysis and applied for the first time to the ground penetrating radar (GPR) data from the tunnel located beneath the Temple of the Feathered Serpent in Teotihuacan, Mexico. The multi-cross wavelet is an extension of the cross wavelet function (Torrence and Compo, 1998), whereas the Fourier multi-cross function is based on the formal definition of the cross function (Wiener, 1928) and is an extension of the Wiener-Khintchine theorem (Khintchine, 1934), both for two or more time series.

Besides their successful use in archeology shown in this paper, these algorithms may be applied to analyze radar data from spaceborne and airborne explorations for water and other resources.

Teotihuacan (Fig. 1) was one of the five most important ancient cities in the world, with around 200,000 inhabitants in an area between 23 and 25 square kilometers. Its main building for a period was the Temple of the Feathered Serpent, located in the Citadel, one of its most important ritual architectural compounds. Discovered and explored by archeologist Manuel Gamio between 1917 and 1922 (Gamio, 1922), the Temple is one of the most impressive and majestic buildings, not just in the ancient city of Teotihuacan but in the American continent. The harmonious integration of its sculpture and architecture, its magnificent details, burials and offerings, and the symbolism of each of its associated elements are a major way to understand the knowledge, power, ideology and calendar of Teotihuacan history and society.

Sergio Gómez-Chávez and Julie Gazzola discovered the tunnel at the end of 2003 while working on the conservation of the temple. The importance of the tunnel lies in being the 'materialization' of the underworld, an outstanding Mesoamerican mythological region, home of telluric forces emanating from deities, where life was constantly created and recreated. The tunnel leads to the center of the temple and has a west-east orientation. The entrance to the underworld was, according to many stories and myths shared by several Mesoamerican cultures, precisely in that direction.

The underworld was an eminently aquatic region. When the inhabitants of Teotihuacan made the cavity which is the entrance to the tunnel, they knew the water table was between 12 and 15 meters below the surface (Fig. 2). This underground conduit was blocked off around 1800 years ago and no one had entered in it since then. The entrance to the tunnel is a few meters away from the platform added to the temple and runs under both the platform and the temple. The entrance is a vertical duct almost five meters wide that descends 14 meters to the beginning of a 103 m long corridor that leads to a series of underground chambers excavated in the rock.

Before starting the excavations, they were planned between 2005 and 2006 on the basis of preliminary analyses of geophysical exploration methods that have been applied to Mexican archeology only recently. Magnetic, electric and GPR methods have been used in projects to locate ceremonial tunnels inside the Moon Pyramid (Arzate et al., 1990) and the Sun Pyramid (Chávez et al., 2001) of Teotihuacan. Geophysical studies have also been performed in closed sites like Templo Mayor in Mexico City (Barba et al., 1997).

GPR is based on the emission of electromagnetic waves and the reception of their reflections occurring at the dielectric constant and electrical conductivity discontinuities of the materials (Davis and Annan, 1989). Researchers appreciate GPR as non-destructive and non-invasive method that provides high-resolution information in real time (Basile et al., 2000).

For GPR data acquisition, the antenna must be dragged on a surface while it transmits and receives electromagnetic waves (monostatic system). After a GPR inspection, a profile of the subsurface strata is obtained in terms of $x$ (distance covered by the antenna) and a two-way travel time (i.e., the time it takes for the electromagnetic pulse to travel from antenna to target and vice versa). The spatial resolution of GPR data depends on the frequency and the sampling frequency used.

Between 2005 and 2006 a GPR equipped with 100 MHz antennas was used to explore the tunnel. Analysis of the GPR traces estimated that the tunnel depth was 12-14 meters and its length was 100-120 meters, with several chambers at its end. Archeologists believe that due to the tunnel's sacredness and importance, one of the chambers may be the tomb of a ruler of the ancient city. Confirmation of this hypothesis would significantly advance knowledge of Teotihuacan history, which lacks evidence about its form of government. Archeological excavations ending in 2014 confirmed the GPR interpretation; the tunnel is 102.43 meters long with three chambers connected at its end.

## 2 Data analysis

*2.1 GPR measurements*

Six measurements ('Tun01'-'Tun06') were carried out during the dry season with GPR pulseEKKO of Sensors & Software, using 100 MHz antennas 1 m apart (bistatic system), on the surface around the tunnel located beneath the Temple of the Feathered Serpent.

'Tun01' was made directly opposite to the entrance of the tunnel pit oriented south to north (19º40'55.62" N Lat., 98º50'49.39" W Long.). 'Tun02' was made at half distance between the pit and the temple (19º40'55.49" N Lat., 98º50'48.89" W Long.). 'Tun03' (Fig. 3) was made at three meters from 'Tun02' (19º40'55.44" N Lat., 98º50'48.75" W Long, whereas 'Tun04' was made at six meters from 'Tun02' (19º40'55.40" N Lat., 98º50'48.63" W Long.). 'Tun03' and 'Tun04' were made exactly over the tunnel. 'Tun05' was made behind the platform, along the alley (19º40'55.02" N Lat., 98º50'47.56" W Long.). 'Tun06' (Fig. 4) was made in the same alley as 'Tun05', but closer to the pyramid of the Feathered Serpent (19º40'55" N Lat., 98º50'47.50" W Long.) In total, 172 traces were obtained for analysis. Several chambers were found at the end of the tunnel (Figs. 5 and 6).

After having determined the interval of the GPR traces inside and outside the tunnel and chamber intervals by means of radargram interpretation of measurements 'Tun03' (near the tunnel, comprising 42 traces along 82 m) and 'Tun06' (near the chamber, comprising 24 traces along 24 m), both multi-cross algorithms described ahead were applied to determine the periods that characterize the matrix, clay and the tunnel and chamber fillings. Figures 3 and 4 show the traces that were analyzed: the trace numbers of traces inside the tunnel/chamber are marked in red, and the trace numbers of those outside the tunnel/chamber are marked in blue.

*2.2 Continuous wavelet transform (CWT)*

Continuous wavelet transform is useful to analyze the evolution of periodicities in the time-frequency domain. It also has the capability to analyze intermittent, non-stationary or transient signals, such as GPR signals. We chose to use the Morlet function as the mother wavelet function because it provides higher resolution in periodicity.

The Morlet wavelet is a complex exponential function modulated by a Gaussian wavelet defined as:

$$\psi(t,s) = e^{i\omega_0 t/s} e^{-t^2/(2s^2)} \qquad (1)$$

Where $t$ is time with $s=(frequency)^{-1}$ as the wavelet scale and $\omega_0$ is a dimensionless frequency.

For our analysis, the wave number is $\omega_0 = 6$ to satisfy the admissibility condition (Farge, 1992). The wavelet power is calculated as:

$$\left|W_n^X\right|^2 \qquad (2)$$

Where $\left|W_n^X\right|$ is the wavelet transform of a time series $X$ and $n$ is the time index (Torrence and Compo, 1998).

The continuous wavelet transform is defined as:

$$W(s,\tau) = \frac{1}{s}\int_{-\infty}^{\infty} X(t)\psi^*\left(\frac{t-\tau}{s}\right)dt \qquad (3)$$

Where $\tau$ is a translation parameter and $s$ refers to the scale.

To calculate the wavelet transform, the time series must have $2^n$ elements. If not, it is padded with zeros and we use the cone of influence (COI), that is the wavelet spectrum region outside of which edge effects become significant (Torrence and Compo, 1998). In this paper, wavelet transforms have a resolution of 0.8 nanoseconds (ns).

## *2.3 Band reconstruction of a time series*

The reconstructed time series is just the sum of the real part of the wavelet transform over all scales (Torrence and Compo, 1998). The decomposition of a signal can be obtained from a time-scale filter. A time-scale filter is defined as:

$$W_n(s) = \frac{\delta_j \delta_t^{1/2}}{C_\delta \psi_0(0)} = \sum_{j=0}^{J} \frac{\text{Re}\{W_n(s_j)\}}{S_j^{1/2}} \qquad (4)$$

Where $\delta_j$ is the factor for scale averaging, $C_\delta$ is a constant ($\delta_j = 0.61$ and $C_\delta = 0.776$, for Morlet wavelet), and $\psi_o(0)$ removes the energy scaling (Torrence and Compo, 1998).

Since the wavelet transform is a band-pass filter with a known response function, it is possible to reconstruct the time series using either deconvolution or inverse filter (Hudgins et al., 1993).

Formally, Parseval's relation for wavelets tells that if $x(t)$ belongs to $L^2(\Re)$, then:

$$\|x(t)\|^2 = \frac{1}{C_\psi} \int ds \int d\tau |W(s,\tau)|^2 \tag{5}$$

The inverse wavelet transform (for almost every $t$) is defined as:

$$x(t) = \frac{1}{C_\psi} \iint W(s,\tau) \psi^* \left(\frac{t-\tau}{s}\right) ds\, d\tau \tag{6}$$

Where $C_\Psi$ is the admissibility constant.

For the reconstruction of the time series analyzed the software programs of Torrence and Compo (1998) were used in this work.

## 2.4 Cross wavelet

The cross wavelet function is defined as:

$$W^{XY} = W^X W^{Y*} \tag{7}$$

Where $X$ and $Y$ represent two different traces and * is the conjugate of the transform $W^Y$.

The cross wavelet indicates where common high power is and reveals information about the phase relationship between the two signals: signals are in phase if the arrow points to the right; they are in antiphase if it points to the left. If the arrow points down, signal one is 90 degrees forward with respect to signal two; and if the arrow points up, signal one is 90 degrees behind signal two.

## 2.5 Multi-cross algorithms

### 2.5.1 Fourier multi-cross function (FMC)

Norbert Wiener (1928), introduced the concept of the cross function which has allowed to find similarities and differences between two time series. This function is defined as:

$$R_{XY}(\tau) = \frac{1}{n}\sum_{i=1}^{n} X(t_i) Y^*(t_i - \tau) \tag{8}$$

With new technologies, researchers tend to use multichannel systems. The new algorithm that we propose in this paper, based on the Hadamard product (Johnson, 1990), is an extension of the Wiener-Khintchine theorem (Khintchine, 1934) that seeks to obtain the cross function of two or more signals.

If there are two time series *X* and *Y* defined as:

$$X(t) = \int G_x(f) e^{ift} df = \mathfrak{I}^{-1}\{G_x(f)\} \tag{9}$$

$$Y^*(t-\tau) = \int G^*_y(f') e^{-if'(t-\tau)} df' \tag{10}$$

The algorithm proposed to apply to GPR measurements is the generalization of the following equation, applied to two time series *X(t)* and *Y(t)*:

$$R_{XY}(\tau) = \int X(t) \cdot Y^*(t-\tau) dt \tag{11}$$

Substituting equations 9 and 10 into 11, we get:

$$R_{XY}(\tau) = \iiint G_x(f) e^{ift} G^*_y(f') e^{-if't} e^{if'\tau} dt df df' = \iiint G_x(f) G^*_y(f') e^{it(f-f')} e^{if'\tau} dt df df' \tag{12}$$

It is known that:

$$\delta(f - f') = \int e^{it(f-f')} dt \tag{13}$$

Substituting 13 into 12, $R_{XY}$ becomes:

$$R_{XY}(\tau) = \iint G_x(f) G^*_y(f') e^{if'\tau} \delta(f - f') df' df \tag{14}$$

Where $\int \delta(f - f') df' = 1$.

Therefore, equation 14 becomes:

$$R_{XY}(\tau) = \int G_x(f) G^*_y(f') e^{if'\tau} df = \mathfrak{I}^{-1}\{G_x(f) G^*_y(f')\} \tag{15}$$

Equation 15 is known as the Wiener-Khintchine theorem (Khintchine, 1934) for the condition $G_x(f) = G_y(f')$.

The multi-cross function we propose is a generalization of the Wiener-Khintchine theorem for N time series.

For the case of several vectors or variables, the algorithm would be as follows:

$$P = [T_1(t), T_2(t), T_3(t), ..., T_n(t)] \tag{16}$$

Where *P* would be the set of traces obtained with GPR.

If $\Im[T_k(t)] = G_k$ is the spectrum for trace "$k$," we define the multi-cross function as the multiple dot product of the set *P* as:

$$R_{T_1,T_2,T_3,\ldots,T_n} = [T_1 \cdot T_2^* \cdot T_3^* \cdot \ldots \cdot T_n^*] \tag{17}$$

Applying the Fourier transform (FT) to equation 17 and using FT properties for the product of functions, the following equation is obtained:

$$G_{T_1,T_2,T_3,\ldots,T_n} = \Im[T_1 \cdot T_2^* \cdot T_3^* \cdot \ldots \cdot T_n^*] = G_1 \cdot G_2^* \cdot G_3^* \cdot \ldots \cdot G_n^* \tag{18}$$

The multi-cross function is derived by applying the inverse transform to equation 18 to obtain the next equation:

$$R_{T_1,T_2,T_3,\ldots,T_n} = \Im^{-1}[G_1 \cdot G_2^* \cdot G_3^* \cdot \ldots \cdot G_n^*] \tag{19}$$

This is the Wiener-Khintchine theorem for the particular case of $G_1 \cdot \ldots \cdot G_n^* = 1$. Equation 19 is the generalization of this theorem for N time series and it is also the algorithm to calculate the cross function by Fourier transform, which from now on will be called Fourier multi-cross function (FMC), obtained for the first time in this work.

### 2.5.2 Multi-cross wavelet (MCW)

Another multi-cross algorithm used in this paper is based on the cross wavelet generalization (Soon et al., 2014):

$$W_{XYZ\ldots n} = W^{X_1,X_2,X_3,\ldots,X_m} = \left\langle W_{F_i} \prod_{k=1}^{m} W_{G_k} \right\rangle \tag{20}$$

Where *F(t)* and *G(t)* are matrices in which each element represents a time-dependent function, $\langle \rangle$ indicates an average of the multi-cross wavelet and $\prod$ a product.

The phase angle of $W^{X_1,X_2,X_3,\ldots,X_m}$ describes the phase relationship between the time series $X_1$, $X_2$, $X_3$, ... and $X_m$ in time-frequency space. Statistical significance of the multi-cross wavelet is estimated by Monte Carlo methods with red noise to determine a 5% significance level (Torrence and Webster, 1999). This function measures the common power between the time series and will be called the multi-cross wavelet (MCW) from now on.

# 3  Results of data analysis

MCW and FMC algorithms were used to analyze the GPR traces of 'Tun03' and 'Tun06' measurements. The spectral colors are in RGB mode: red indicates the highest spectral power and blue the lowest spectral power. In all figures the panel above the wavelet spectrum shows the time series (traces). The lower panel shows the instantaneous phase which describes the type of correlation between the time series: linear when it is close to zero and complex otherwise. Periods and scales (right panel next to the wavelet spectrum) are in ns.

## 3.1 'Tun03' measurement

Figures 7-10 correspond to the analysis of the following tunnel intervals.

### 3.1.1  Spectra for traces inside-and-outside the tunnel interval

Multi-cross wavelet spectra were obtained for GPR data of traces inside (traces 21-23) and outside the tunnel interval (traces 4-5). In MCW analysis the 15.22 ns period corresponded to the dominant period (Fig.7).

The dominant period of the FMC spectrum was 9.05±3.35 ns (Fig. 8).

### 3.1.2  Spectra for traces outside the tunnel interval

Multi-cross wavelet spectra were obtained for GPR data outside the tunnel interval (traces 1-5). In MCW analysis the 20.32 ns period corresponded to the dominant period (Fig. 9A).

The FMC spectrum was filtered with a 6-10 ns band because in this interval a dominant period of 5.70 ns was observed (Fig. 9B). The dominant period of the filtered FMC spectrum was 7.18±0.78 ns (Fig. 9C).

### 3.1.3  Spectra for traces inside the tunnel interval

GPR data inside the tunnel interval were those from traces 19-23. In MCW analysis, the 14.37 ns period corresponded to the dominant period (Fig. 10A).

The FMC spectrum was filtered with a 2-5 ns band because in this interval several reflections were observed between periods 1.80 and 2.85 ns (Fig. 10B). The dominant period of the filtered spectrum was 5.08±1.08 ns (Fig. 10C).

## 3.2 'Tun06' measurement

Figures 11-14 correspond to the following chamber intervals.

### 3.2.1 Spectra for traces inside-and-outside the chamber interval

Multi-cross wavelet spectra were obtained from traces inside (traces 14 and 15) and outside the chamber interval (traces 20-22).

The dominant periods of the MCW and FMC spectra were 11.40±1.00 ns (Fig. 11) and 10.16±1.76 ns (Fig. 12), respectively.

### 3.2.2 Spectra for traces outside the chamber interval

GPR data outside the chamber interval were those from traces 20-24. In MCW analysis the 20.32±0.68 ns period corresponded to the dominant period (Fig. 13A).

The FMC spectrum was filtered with a band from 6 to 10 ns because in this interval a dominant period of 9.05 ns was observed (Fig. 13B). The dominant period of the filtered FMC spectrum was 8.06±1.06 ns (Fig. 13C).

### 3.2.3 Spectra for traces inside the chamber interval

GPR data inside the chamber interval were those from traces 14 and 15. In MCW analysis the 11.40±0.40 ns period corresponded to the dominant period (Fig. 14A).

The FMC spectrum was filtered with a 2-5 ns band because in this interval a high power period of 3.59 ns was observed (Fig. 14B). The dominant period of the filtered FMC spectrum was 4.27±0.82 ns (Fig. 14C).

The periods of three intervals —traces inside-and-outside the tunnel/chamber, outside the tunnel/chamber and inside the tunnel/chamber— analyzed with the MCW and filtered FMC algorithms are summarized in Table 1.

## Discussion

Two multi-cross algorithms, MCW and FMC, were used in this work to analyze GPR data in order to determine the periods corresponding to subsurface strata of the tunnel beneath the Temple of the Feathered Serpent. The MCW algorithm had already proved to be useful in the analysis of solar activity (Soon et al., 2014). The FMC algorithm is based on the Hadamard product (Johnson, 1990) and the formal definition of the cross function (Eq. 7). It is an extension of the Wiener-Khintchine theorem for two or more time series. This algorithm is described here for the first time and also applied with MCW to analyze GPR data.

Analysis of three intervals —i.e., GPR traces inside-and-outside the tunnel/chamber, outside the tunnel/chamber and inside the tunnel/chamber— with both the MCW and the filtered FMC algorithms led us to determine the periods corresponding to the tunnel and chamber fillings, as well as to the clay and the matrix.

The period of the tunnel filling obtained by MCW analysis (14.37 ns) reflects the mixed limestone and clay compound contained in this stratum since its value is close to that of the period of the matrix (15.22 ns). Analyzed in the same way, the periods of the chamber filling (11.40±0.40 ns) and that of the matrix (11.40±1.00 ns) were almost identical, as well as the clay periods from 'Tun03' (20.32 ns) and 'Tun06' (20.32±0.68 ns).

FMC analysis of the tunnel obtained a period (5.08±1.08 ns) close to that of the chamber (4.27±0.82 ns), suggesting that the tunnel and chamber are filled with similar materials. Using the same algorithm, the clay periods were also similar for 'Tun03' (7.18±0.78 ns) and 'Tun06' (8.06±1.06 ns), as well as those of the matrix for 'Tun03' (9.05±3.35 ns) and 'Tun06' (10.16±1.76 ns).

Comparing the results of Figs. 10 and 13, it can be observed that the MCW analysis limits the spectral power satisfactorily and that the main periodicities of the site analyzed can be obtained. By FMC analysis, the spectral power cannot be located in time with precision. In order to solve this problem, its spectrum must be filtered.

The results obtained by FMC analysis could be of interest for GPR equipment that has the Fourier transform analysis tool. The FMC algorithm described in Section 2.5.1 could be implemented in the processing of GPR data for equipment that does not have the wavelet analysis tool.

A methodology for the FMC algorithm is proposed as follows:

a) Obtain the product of GPR traces, according to Section 2.5.1.
b) Perform the Fourier spectral analysis of the product of the traces to find the main periodicities.
c) Perform a low pass or high pass filter.
d) Compare the result with MCW analysis. This last may be omitted when the user gains experience in step c).

Idi and Kamarudin (2012), Ni *et al.* (2010), Jeng *et al.* (2009) and Baili *et al.* (2006) use the continuous wavelet concept for multiresolution radargram image analysis. In contrast, we show here how raw GPR data can be used to characterize strata instead of only radargram images. Noise can be filtered with time series reconstruction, and two or more traces can be compared using the two multi-cross algorithms.

The use of both algorithms allows a deeper analysis since the similarities of the tunnel and chamber filling periods could not have been determined with the MCW algorithm alone.

Our results do not imply that radargram interpretation should be discarded because our algorithms are not used to detect but to characterize the strata of interest that have been identified by the radargram analyses.

The successful application of the multi-cross algorithms shown here for archeological GPR data suggests its use may be extended to the search of water and other resources in celestial bodies.

## Acknowledgements

FLR was supported by a scholarship from CEP-UNAM. VMVH acknowledges financial support from the CONACyT-180148 grant. SGC and JG acknowledge support from Instituto Nacional de Antropología e Historia (Mexico). This work is a contribution to the TLALOCAN project.

**Table 1**

**Values of the periods (ns) obtained from the strata analyzed with the MCW, FMC and filtered FMC algorithms**

| GPR measurement | Algorithm | Intervals | | | |
|---|---|---|---|---|---|
| | | Inside-and-outside the tunnel/chamber | Outside the tunnel/chamber | Inside the tunnel | Inside the chamber |
| Tun03 | MCW | 15.22 | 20.32 | 14.37 | NA |
| | FMC | 9.05±3.35 | NA | NA | NA |
| | Filtered FMC | NA | 7.18±0.78 | 5.08±1.08 | NA |
| Tun06 | MCW | 11.40±1.00 | 20.32±0.68 | NA | 11.40±0.40 |
| | FMC | 10.16±1.76 | NA | NA | NA |
| | Filtered FMC | NA | 8.06±1.06 | NA | 4.27±0.82 |

NA, non applicable.

# Figures

**Fig. 1. Location of the tunnel beneath the Temple of the Feathered Serpent in Teotihuacan, Mexico.**

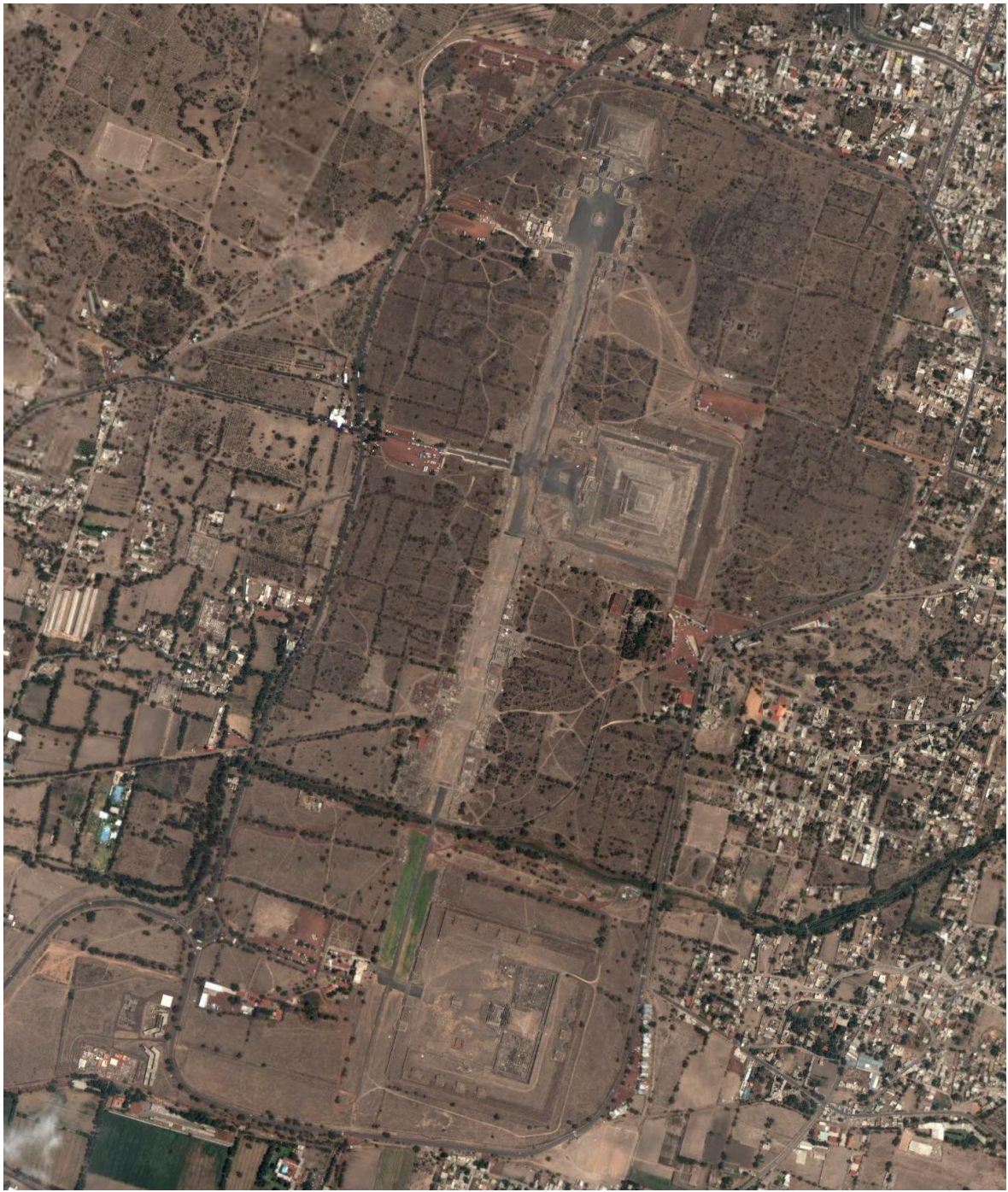

**Fig. 2. First 30 meters of the tunnel beneath the Temple of the Feathered Serpent in Teotihuacan.**

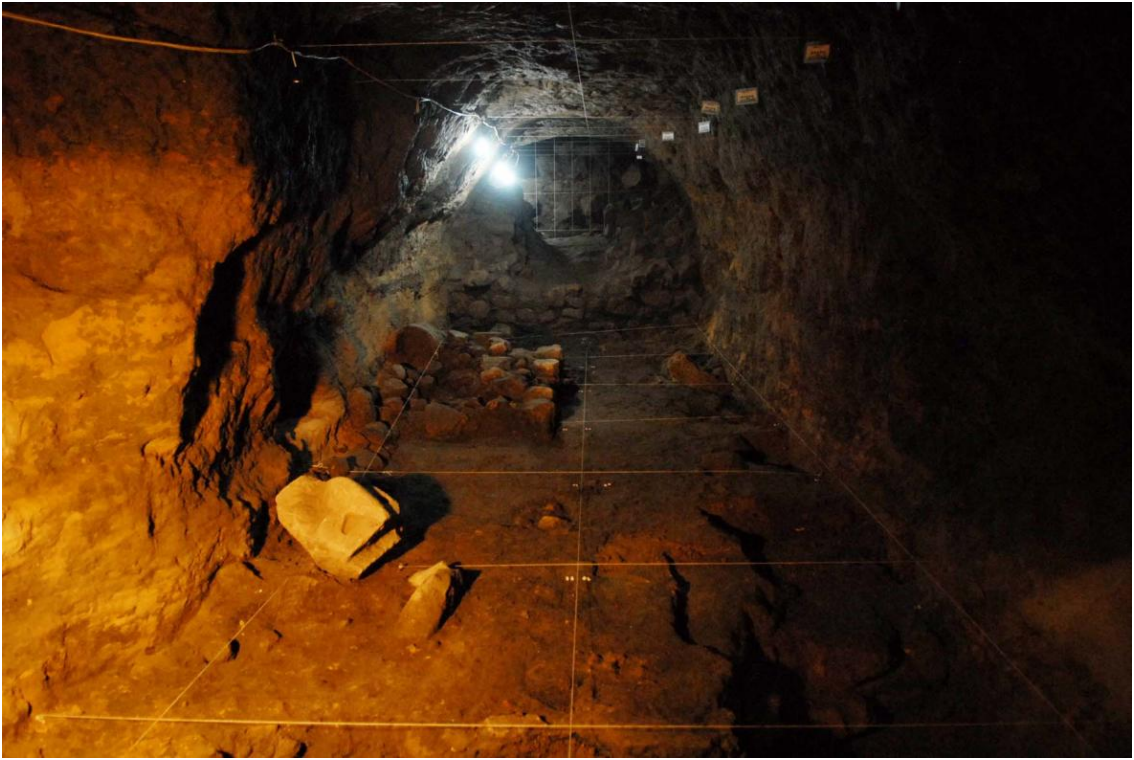

**Fig. 3. Traces of 'Tun03' measurement.** (A) Location. (B) Radargram. Traces inside the tunnel interval are marked in red and traces outside are marked in blue.

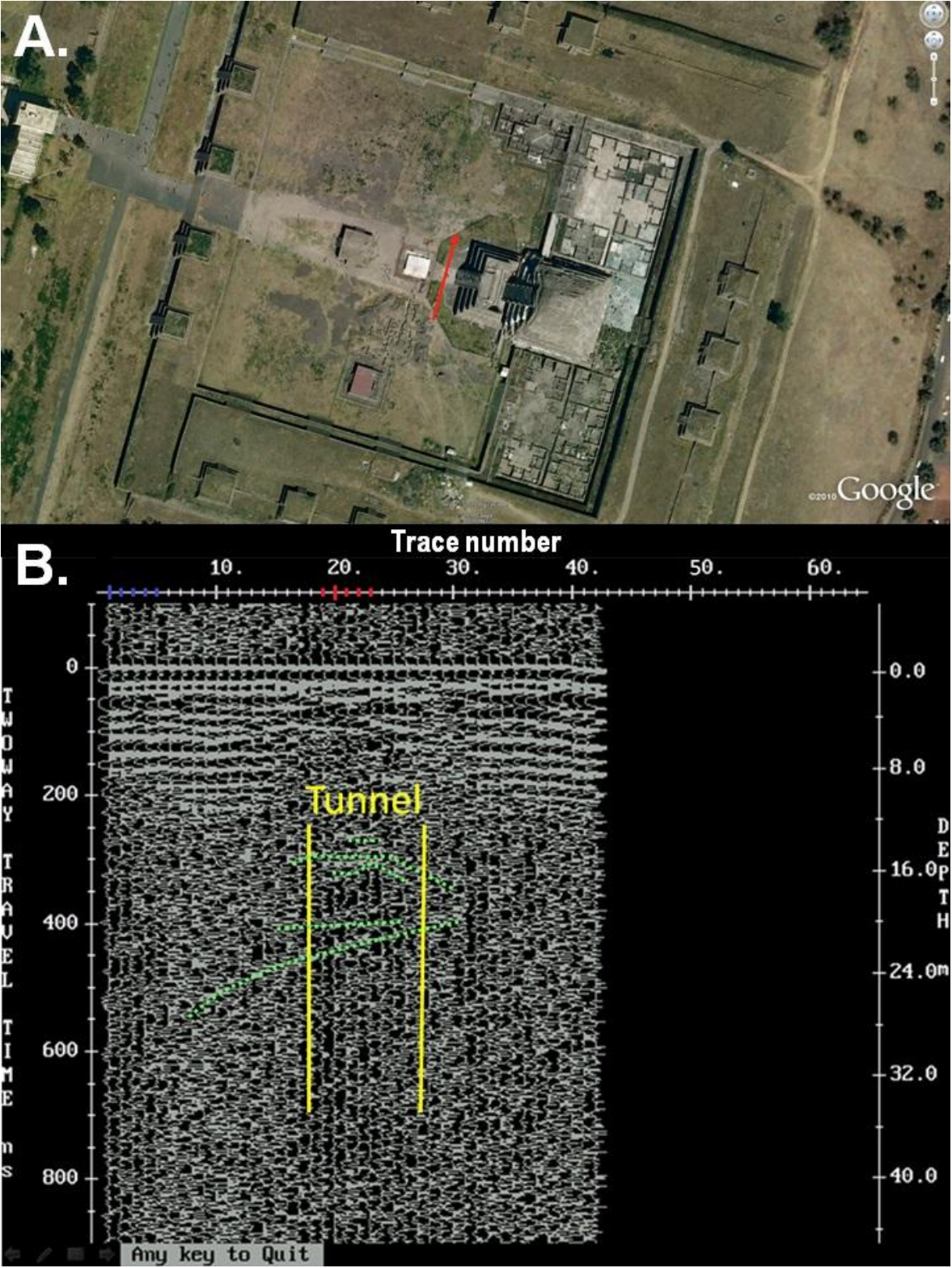

**Fig. 4. Traces of 'Tun06' measurement.** (A) Location. (B) Radargram. Traces inside the chamber interval are marked in red and traces outside are marked in blue.

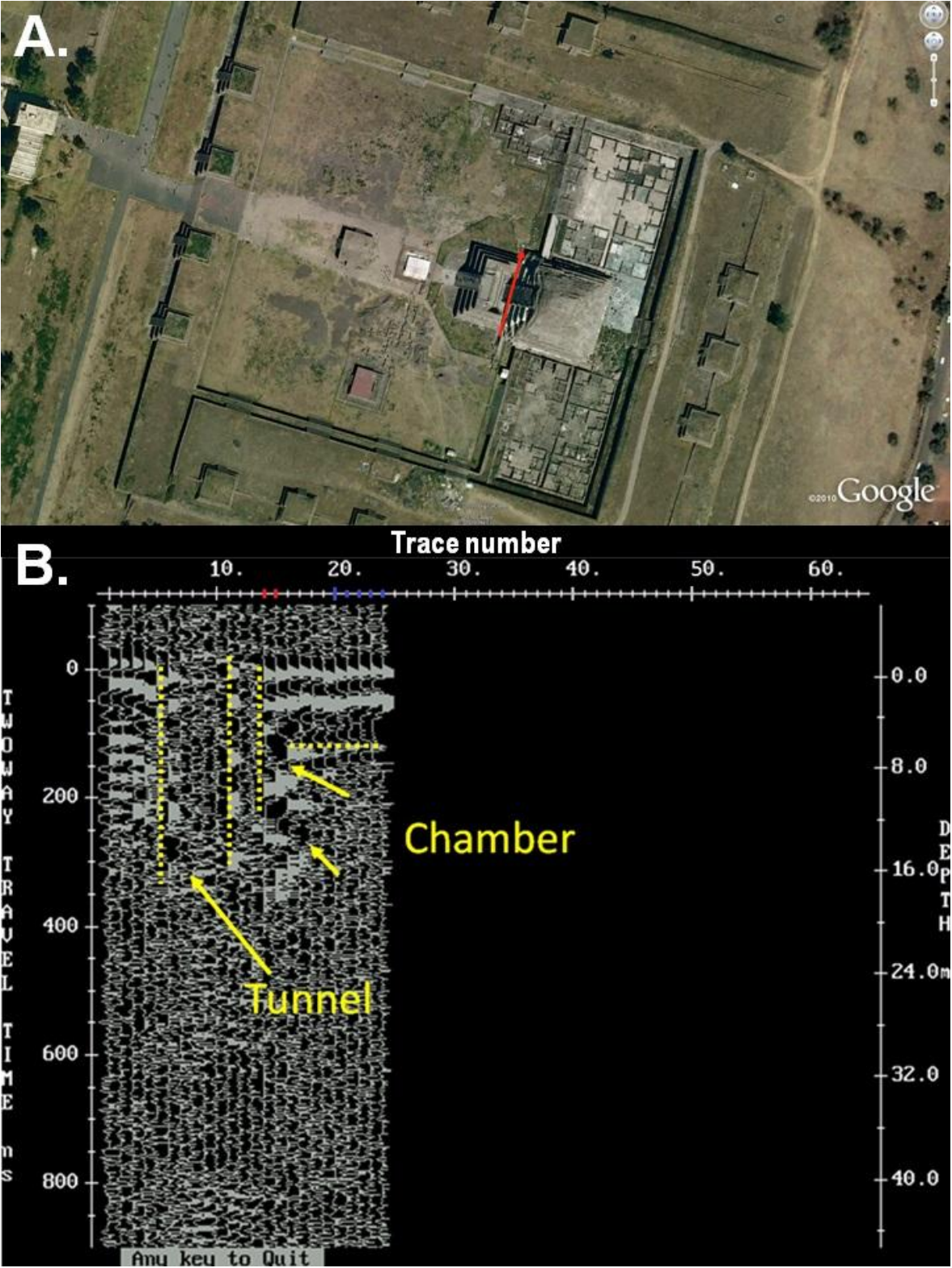

**Fig. 5. View inside the tunnel beneath the Temple of the Feathered Serpent.**

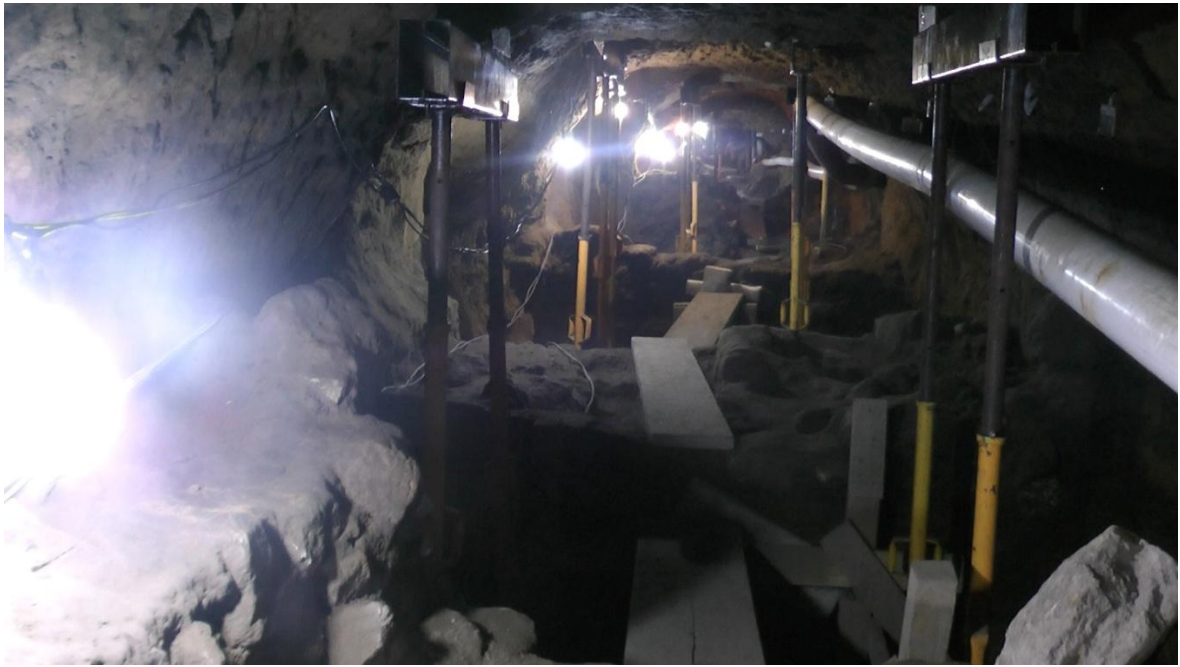

**Fig. 6. View inside one of the chambers at the end of the tunnel.**

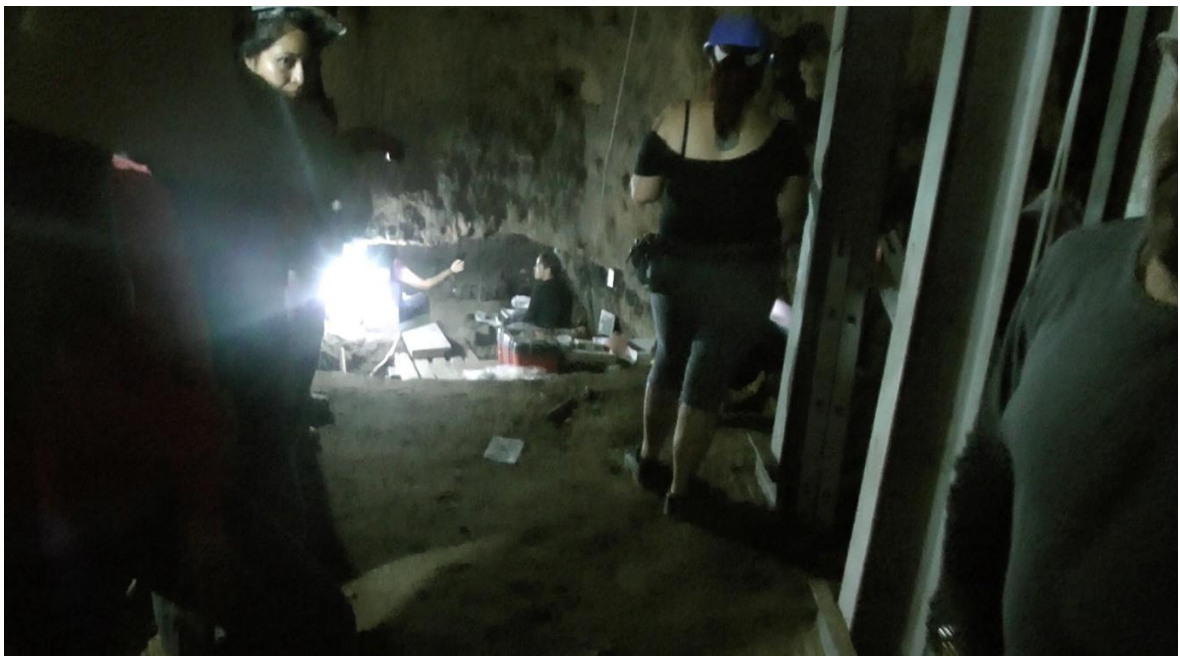

**Fig. 7.** MCW spectrum of traces 19-21 (inside the tunnel) and of traces 4-5 (outside the tunnel) from 'Tun03' measurement.

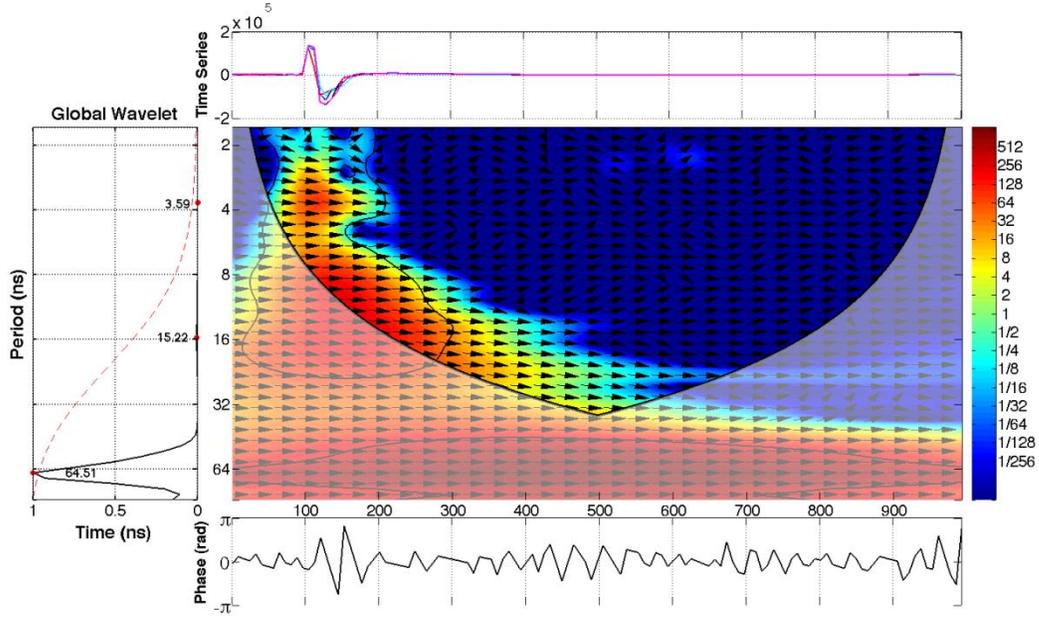

**Fig. 8.** FMC wavelet spectrum of traces 19-21 (inside the tunnel) and of traces 4-5 (outside the tunnel) from 'Tun03' measurement.

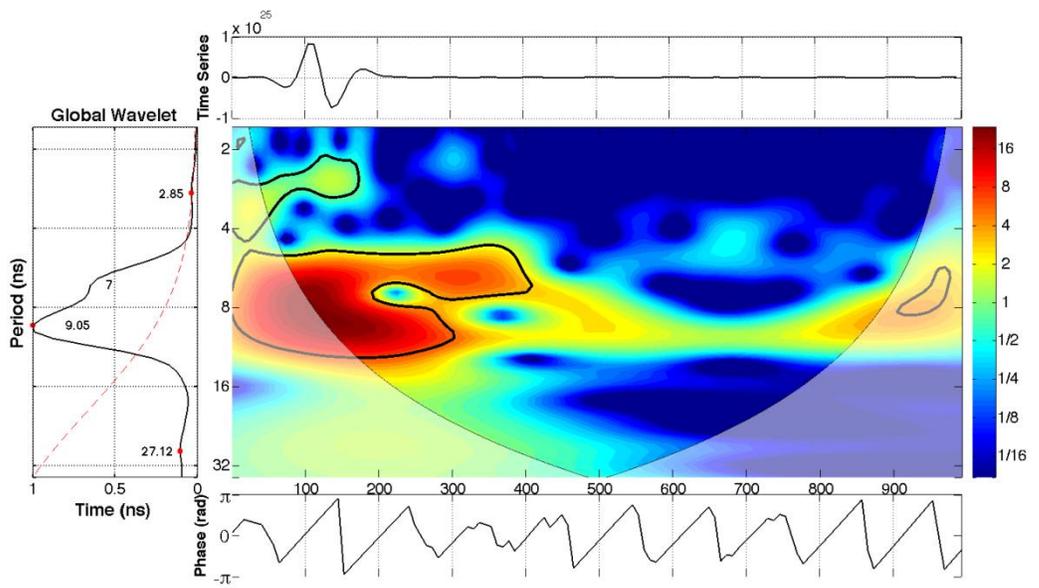

**Fig. 9. Multi-cross wavelet spectra of traces 1-5 (outside the tunnel) from 'Tun03' measurement.** (A) MCW. (B) FMC. (C) Filtered FMC (band 6-10 ns).

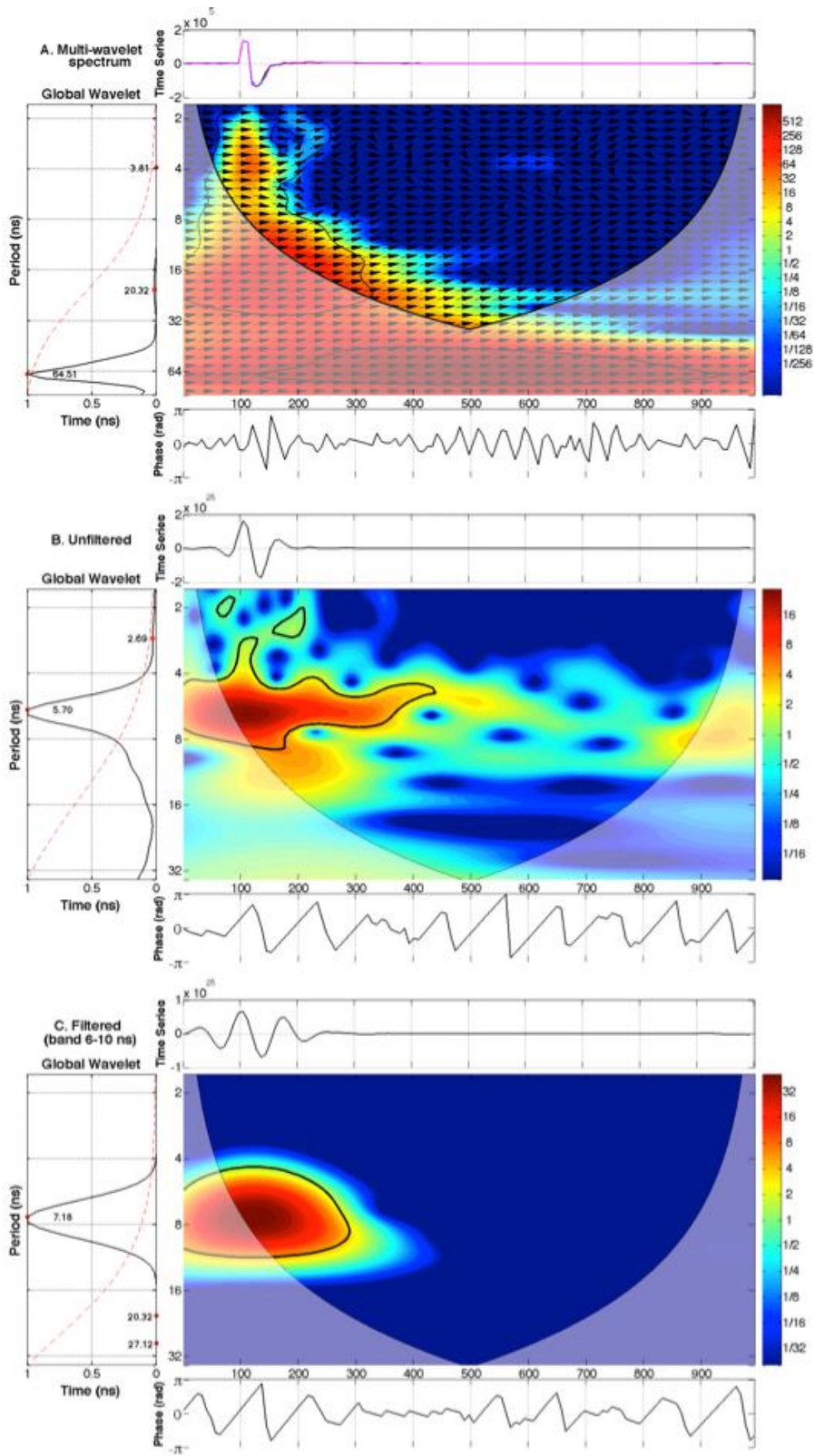

**Fig. 10. Multi-cross wavelet spectra of traces 19-23 (inside the tunnel) from 'Tun03' measurement.** (A) MCW. (B) FMC. (C) Filtered FMC (band 2-5 ns).

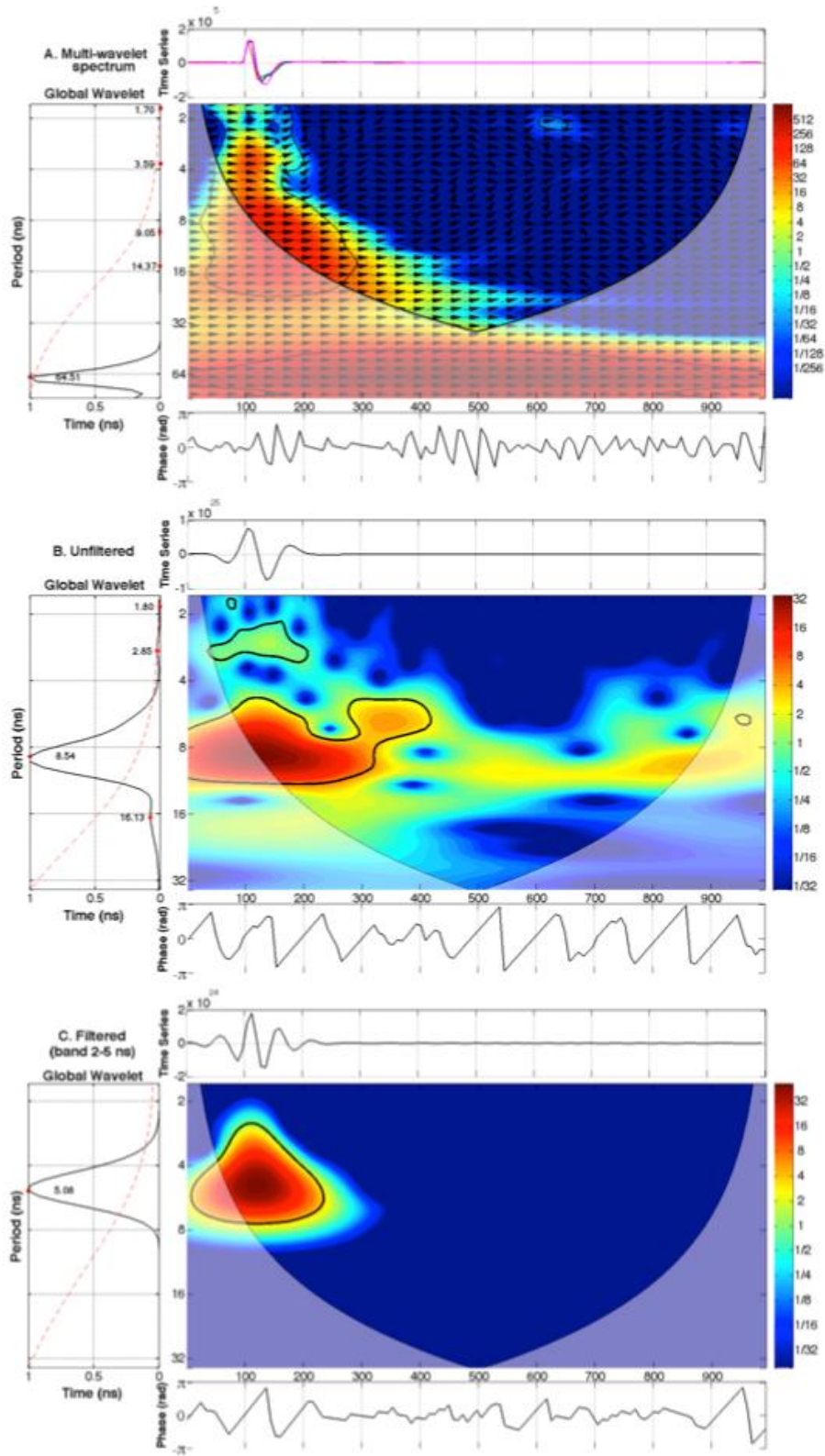

**Fig. 11. MCW spectrum of traces 14-15 (inside the chamber) and of traces 20-22 (outside the chamber) from 'Tun06' measurement.**

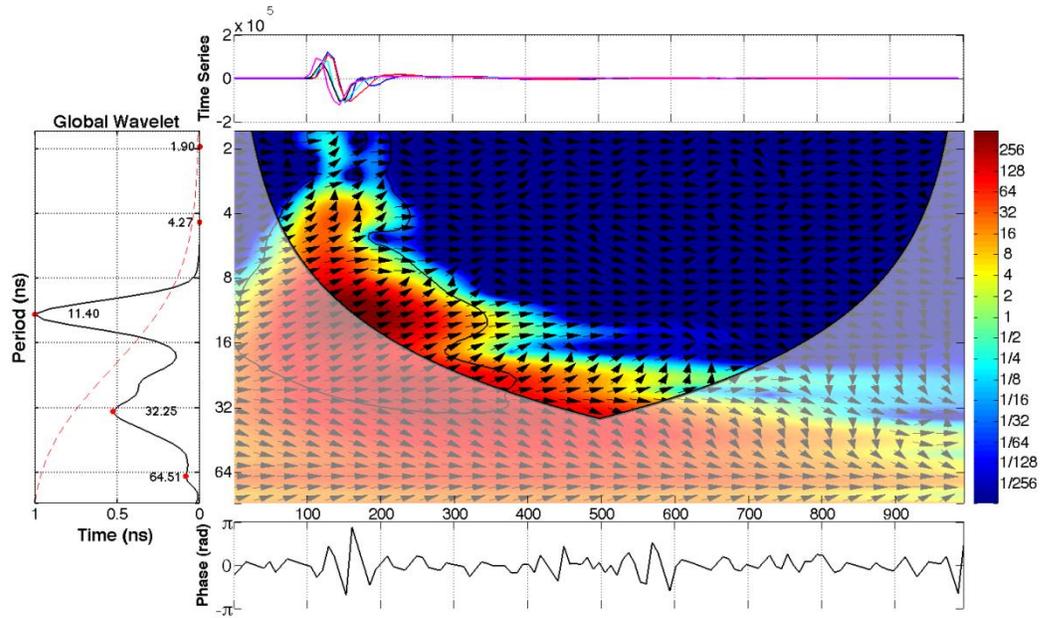

**Fig. 12. FMC wavelet spectrum of traces 14-15 (inside the chamber) and of traces 20-22 (outside the chamber) from 'Tun06' measurement.**

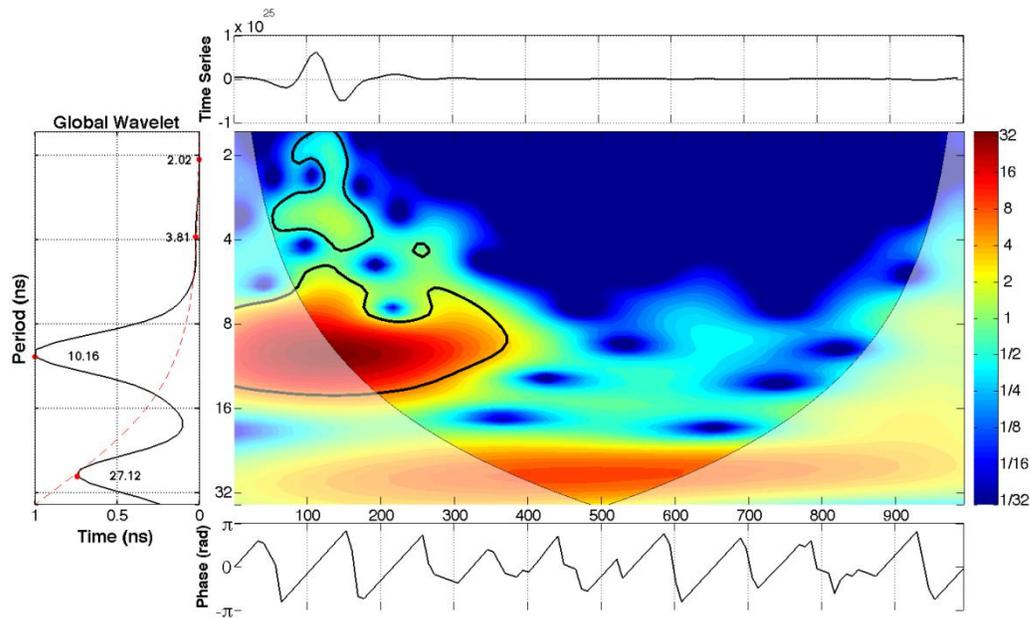

**Fig. 13. Multi-cross wavelet spectra of traces 20-24 (outside the chamber) from 'Tun06' measurement.** (A) MCW. (B) FMC. (C) Filtered FMC (band 6-10 ns).

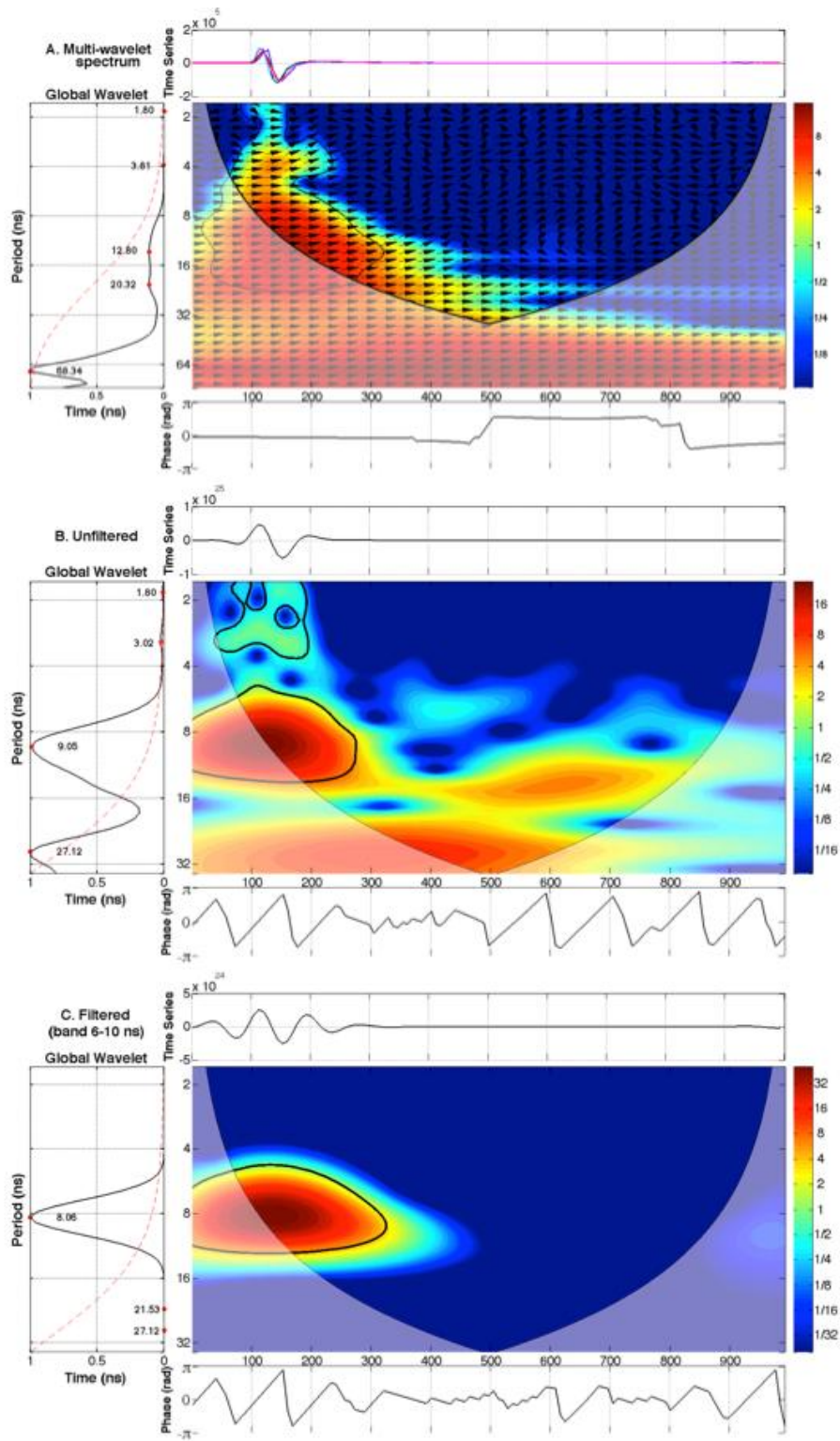

**Fig. 14. Multi-cross wavelet spectra of traces 14 and 15 (inside the chamber) from 'Tun06' measurement.** (A) MCW. (B) FMC. (C) Filtered FMC (band 2-5 ns).

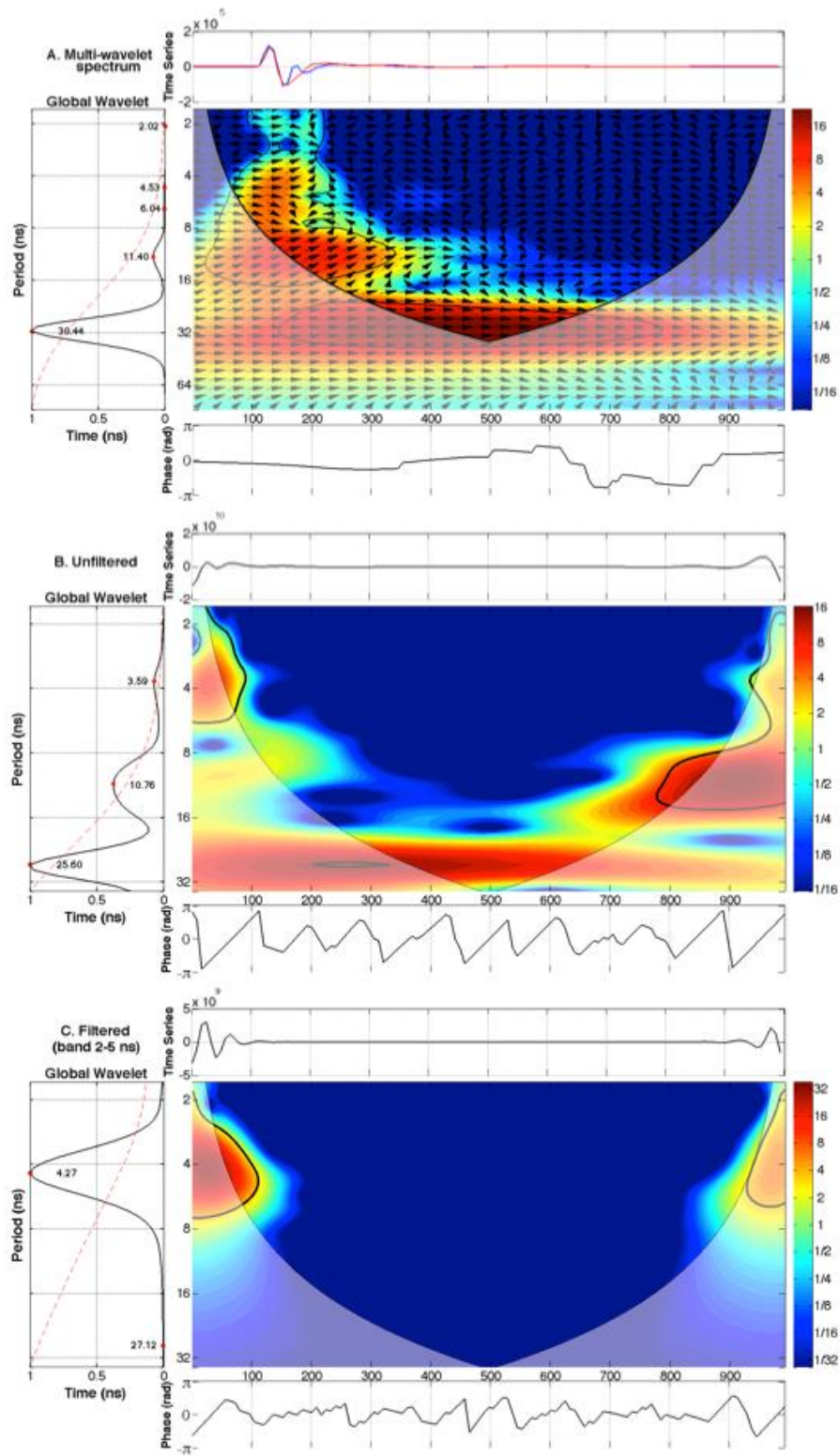